\documentclass{osa-article}

\journal{oe}

\articletype{Research Article}

\usepackage{lineno}
\usepackage{graphicx}
\usepackage{dcolumn}
\usepackage{bm}

\usepackage[utf8]{inputenc}
\usepackage[T1]{fontenc}
\usepackage{mathptmx}
\usepackage{etoolbox}

\usepackage{xcolor}

\newcommand{\avdw}{\alpha_{\mathrm{vdW}}}

\begin{document}

\title{Cryogenic and hermetically sealed packaging of photonic chips for optomechanics}

\author{W. W. Wasserman\authormark{1,2,*}, R. A. Harrison\authormark{1,2},  G. I. Harris\authormark{1}, A. Sawadsky\authormark{1}, Y. L. Sfendla\authormark{1}, W. P. Bowen\authormark{1} and C. G. Baker\authormark{1}}

\address{\authormark{1}ARC Centre of Excellence for Engineered Quantum Systems, School of Mathematics and Physics, University of Queensland, St Lucia, QLD 4072, Australia.}
\address{\authormark{2} These authors contributed equally}
\email{\authormark{*}w.wasserman@uq.edu.au}

\begin{abstract}
We demonstrate a hermetically sealed packaging system for integrated photonic devices at cryogenic temperatures with \emph{plug-and-play} functionality. This approach provides the ability to encapsulate a controlled amount of gas into the optical package allowing helium to be used as a heat-exchange gas to thermalize photonic devices, or condensed into a superfluid covering the device. This packaging system was tested using a silicon-on-insulator slot waveguide resonator which fills with superfluid $^4$He below the transition temperature. To optimize the fiber-to-chip optical integration 690 tests were performed by thermally cycling  optical fibers bonded to various common photonic chip substrates (silicon, silicon oxide and HSQ) with a range of glues (NOA 61, NOA 68, NOA 88, NOA 86H and superglue). This showed that NOA 86H (a UV curing optical adhesive with a latent heat catalyst) provided the best performance under cryogenic conditions for all the substrates tested. The technique is relevant to superfluid optomechanics experiments, as well as quantum photonics and quantum optomechanics applications.
\end{abstract}

\section{\label{sec:level1}Introduction} 
Cryogenic photonic devices have a variety of applications, ranging from quantum photonics~\cite{elshaari_hybrid_2020} (where devices such as single quantum  emitters~\cite{somaschi_near-optimal_2016,singh_quantum_2019} and superconducting single photon detectors~\cite{sprengers_waveguide_2011} require cryogenic temperatures) to quantum optomechanics~\cite{verhagen_quantum-coherent_2012,riedinger_remote_2018,shomroni_optical_2019} and the study of the exotic properties of superfluid helium\cite{harris_laser_2016, sfendla_extreme_2021,sachkou_coherent_2019}. Here we detail a procedure for integrating photonic devices for cryogenic applications that is robust and reliable without the need for \textit{in situ} optical alignment, as well as a system for packaging these devices so they can be easily integrated into any cryostat with \emph{plug-and-play} usability via an optical fiber connector. A key novelty here is the ability to encapsulate a controlled amount of gas into the optical package. This allows for  exchange-gas  thermalization of photonic devices when used with helium gas~\cite{verhagen_quantum-coherent_2012,riviere_evanescent_2013,macdonald_optomechanics_2016,shomroni_optical_2019}. Condensation of thin films on the photonic device also allows for a broad range of other experiments~\cite{srinivasan_optical_2007}.

A technique for the optical coupling to photonic chips at cryogenic temperatures using grating couplers and angle-polished fibers has recently been reported by McKenna et al.~\cite{mckenna_cryogenic_2019}, which we build upon here. We find that the success of the method is strongly dependent on the choice of substrate material and cryogenic adhesive. For this reason, we perform an in-depth analysis of the bonding strength with five optical adhesives and three substrate materials, including a study of the survival rate over multiple thermal cycles. The outcomes of this enable the efficient bonding to grating couplers with and without an oxide or hydrogen silsequioxane (HSQ) cladding. We extend the technique to adiabatic fiber couplers~\cite{groblacher_highly_2013,tiecke_efficient_2015} for broadband alignment-free cryogenic optical coupling. 

As a proof of principle experiment, we package a silicon-on-insulator slotted ring resonator. The resonator is evanescently coupled to a waveguide which is coupled to optical fiber via grating couplers. The device is then hermetically sealed in a helium-4 gas environment. Using helium as a buffer gas~\cite{riviere_evanescent_2013}---particularly helium-3~\cite{verhagen_quantum-coherent_2012, macdonald_optomechanics_2016,shomroni_optical_2019}--- enables efficient device thermalization at cryogenic temperatures due to its high thermal conductivity. This is particularly useful for nanoscale optomechanical devices such as optomechanical crystals~\cite{meenehan_pulsed_2015,riedinger_remote_2018,mckenna_cryogenic_2019,shomroni_optical_2019}, where the small optical mode volumes  and reduced thermal anchoring  required to limit acoustic losses lead otherwise to prohibitive optical absorption heating. This typically limits experiments to low photon numbers or pulsed operation~\cite{meenehan_pulsed_2015,riedinger_remote_2018}. In contrast, the use of $^3$He buffer gas has recently enabled the strong continuous laser drives required for backaction-evading measurements~\cite{shomroni_optical_2019} and laser cooling such nanomechanical resonators to their zero-point energy~\cite{qiu_laser_2020}. Condensed as a superfluid,  helium-4 may interact with the photonic chip for a range of applications, including cavity optomechanics and Brillouin scattering~\cite{lorenzo_superfluid_2014,harris_laser_2016,kashkanova_superfluid_2017,he_strong_2020,harris_proposal_2020,spence_superfluid_2021},  quantum information~\cite{koolstra_coupling_2019}, the study of quantum hydrodynamics and turbulence \cite{sachkou_coherent_2019}, as well as applications in gravitational wave~\cite{vadakkumbatt_prototype_2021} or dark matter detection~\cite{schutz_detectability_2016}. In our experiments we observe the condensation of superfluid helium-4 films a few nanometers thick on the ring resonator via the optical frequency shift this generates---confirming that the quantity of enclosed gas within our package can be precisely controlled---and observe superfluid acoustic waves confined to the device.

\section{Methods}
\subsection{Requirements}
 
It is broadly challenging to perform cryogenic experiments with integrated photonics, especially where an exchange gas is required. Technical requirements may include optical fiber feedthroughs, which need to be vacuum-tight, and in the case of helium, superfluid tight~\cite{butterworth_demountable_1998}.

\begin{figure}
\centering
\includegraphics[width=\textwidth]{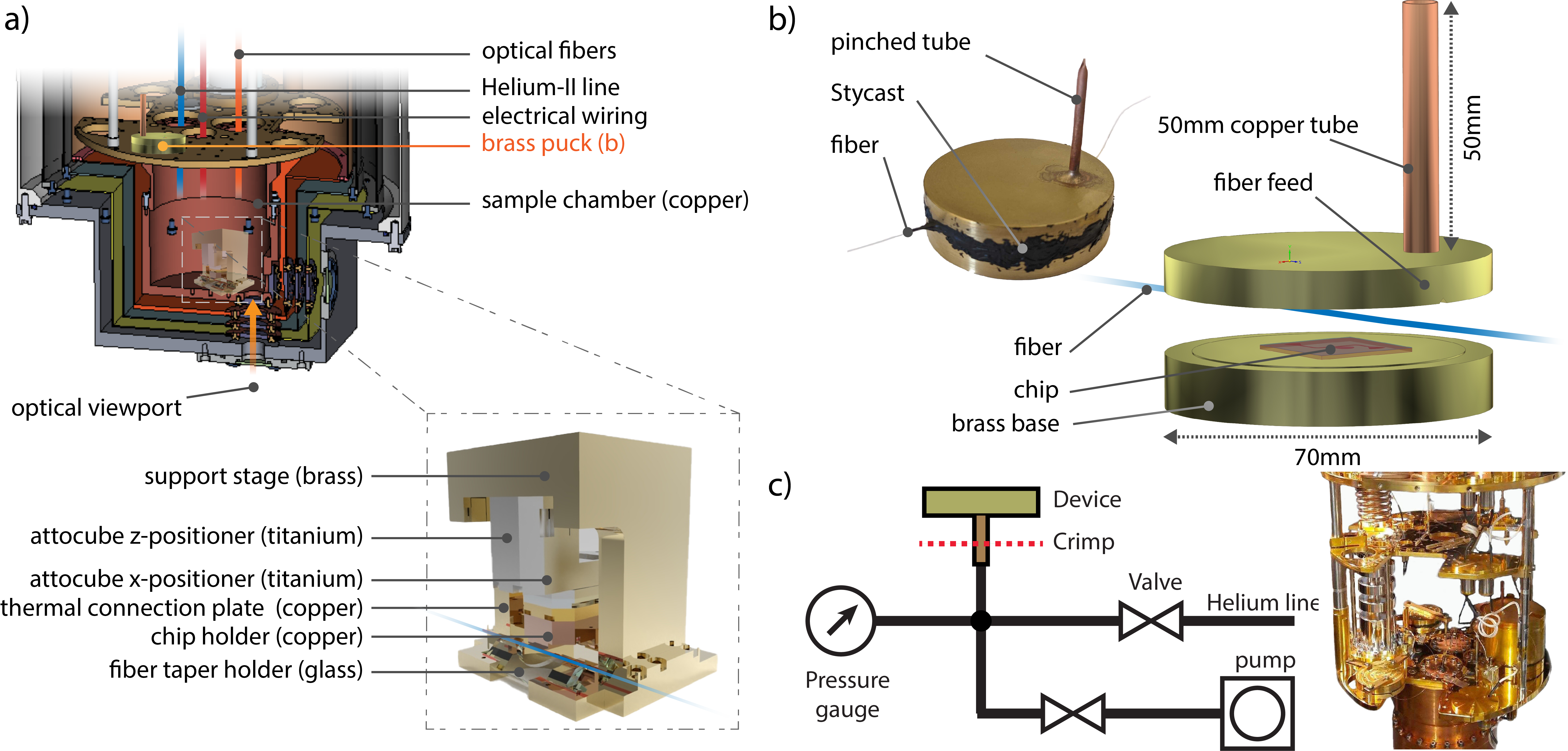}
\caption{a) Complex experimental apparatus for superfluid optomechanics experiment, where light is coupled into the chip via a fiber taper. For alignment purposes an optical viewport and positioning stages are required. The brass puck is shown for size comparison. b) Left: image of a packaged puck with a photonic device inside ready to be installed in a cryostat. Right:  CAD rendering of the brass puck, showing a photonic chip and device dimensions. (c) Left: Schematic of the gas filling process. Right: image of three `puck' devices on the mixing chamber flange of a dilution refrigerator, marked by the grey arrows.}
\label{fig:PuckSchematic}
\end{figure}

Gases are typically injected during cryogenic operation \cite{kashkanova_optomechanics_2018,yang2016coupling,he_strong_2020}, requiring cryogenic feedthroughs for the cryostat. A typical schematic for a setup such as employed in reference \cite{he_strong_2020} is shown in Fig.~\ref{fig:PuckSchematic}~(a). In addition to the capillary and superfluid-tight hermetic seals, the cryogenic alignment and evanescent coupling of a fragile optical fiber taper to a device requires electrical feedthroughs for nanopositioners, as well as visual access achieved through five windows in the vacuum can and thermal shields of the dilution refrigerator. This complexity adds to the thermal challenges of running experiments at cryogenic temperatures.

Here a greatly simplified \emph{plug-and-play} approach requiring only fiber access is demonstrated. This is shown in Fig.~\ref{fig:PuckSchematic}~(b), and consists of a hermetically sealed cylindrical brass container hosting an integrated photonic chip fiber-coupled via grating couplers through a cryogenically robust gluing protocol. The difficulty of using integrated photonics in this cryogenic environment is alleviated by developing a robust gluing procedure of the optical fibers that couple light into the chip. Once on the chip, light is coupled into optical components through monolithically integrated waveguides which do not require \textit{in situ} alignment.  Bonding fibers to the chip thus removes the requirement for coupling to devices with a tapered fiber and the associated electronic nanopositioners and imaging system. This  gas and superfluid-tight vessel even eliminates the requirement for a  capillary, being prepackaged with the precise amount of gas required for the experiment. This greatly reduces the barrier to entry for superfluid experimentation through the use of a prepackaged optical component that can be mounted to a cold stage of a standard cryostat as shown in Fig.~\ref{fig:PuckSchematic}~(a).  

\subsection{Packaging}

The `puck' shown in Fig.~\ref{fig:PuckSchematic}~(b) on the left shows an image of a final device on the left with plenty of epoxy used to seal the two brass sections together. The CAD drawing in Fig.~\ref{fig:PuckSchematic}~(b) on the right illustrates a photonic chip with devices to be attached inside the puck. Once the optical fibers have been connected, as described in section \ref{sectiongluing}, the photonic chip can be secured to the brass base as preferred. Here a thin layer of \emph{Apiezon N} cryogenic vacuum grease is used to hold the chip in place and provide thermal contact between the chip and brass. Stycast epoxy (2850FT) with catalyst type 9 is then prepared and applied in a thin layer to both faces of the 5~mm wide sealing section, carefully ensuring that the v-shaped fiber access grooves are filled (see Fig.~\ref{fig:PuckSchematic}~(b)). This epoxy was chosen as its thermal expansion is similar to copper \cite{ventura_thermal_2014}. The v-shaped grooves are aligned to the optical fibers and the two sides of the puck are firmly pressed together. The Stycast that gets pressed out is then smeared across the joint to ensure that no gaps remain.  After 24 hours the Stycast is sufficiently cured to add helium via the pinch-off tube (copper CDA 101, 0.125" diameter: Solid Sealing Technologies FA202XFXFC030) silver soldered to the brass puck. The use of this small diameter oxygen free copper (degreased and deoxidized with acetone and citric acid) is necessary for creating a superfluid tight hermetic seal with a leak rate measured to be below $10^{-10}$~mbar$\cdot$L/s. The container is evacuated using a roughing pump before being filled with the desired gas pressure, as shown in Fig.~\ref{fig:PuckSchematic}~(c). Once filled, the pinch-off tube is hermetically sealed using a hand operated hydraulic crimp tool (Solid Sealing Technologies KT35046), which forms a superfluid-tight cold weld (dashed red line in Fig.~\ref{fig:PuckSchematic}~(c)).

\subsection{Optical Integration}
 
\begin{figure}
\centering
    \includegraphics[width=\textwidth]{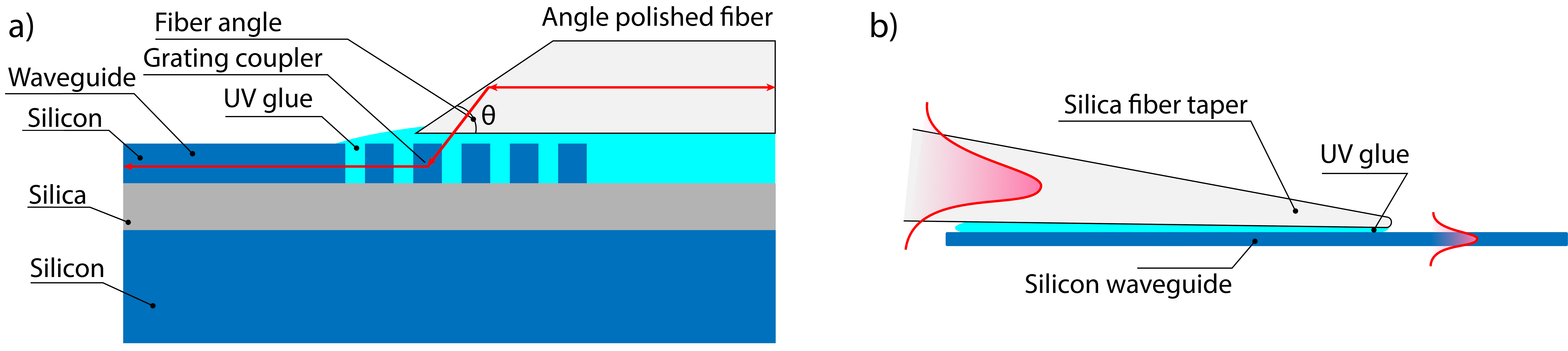}
\caption{(a) Side view sketch of a  $\theta=35^\circ$ angle-polished fiber reflecting light down onto a grating coupler, which preferentially scatters the light into a waveguide. The grating coupler and waveguide are made of silicon defined by a lithography process on a silica insulating layer. The glue normally wicks into the grating coupler, but occasionally does not, which is apparent by the lack of optical shift when uncladded. (b) Side view sketch of the adiabatic fiber coupling scheme~\cite{groblacher_highly_2013,tiecke_efficient_2015}. The silicon waveguide is 220~nm thick, fully released and tapered along its transverse direction from a width of 200~nm to 500~nm.}
\label{fig:PhotonicDevice}
\end{figure}

The optical coupling into a device can be done through a variety of methods. Here we demonstrate two methods which are sketched in Fig.~\ref{fig:PhotonicDevice}~(a) and (b): grating couplers with glued angle polished fibers \cite{mckenna_cryogenic_2019} and adiabatic fiber couplers~\cite{groblacher_highly_2013,tiecke_efficient_2015}. The procedure from McKenna et al. \cite{mckenna_cryogenic_2019} was followed in the first case, aligning the angled polished fibers while measuring the optical signal with a laser to optimise the coupling, before fixing these in place with a fast curing UV epoxy (discussed in more detail below). We employ a similar approach in the second case, using a silica taper made from SMF-28 fiber pulled under a hydrogen flame. The taper is dipped in UV epoxy and the excess wiped off on the edge of the chip before performing alignment and UV curing. In this second method, successful cryogenic operation was only achieved through the use of a UV epoxy containing a latent heat catalyst (NOA 86H), which we describe below. Using these techniques we achieve coupling efficiencies and bandwidths at cryogenic temperatures of $\sim 10$~\% and 20~nm for the grating coupler, and $\sim30$~\% and $\gg$ 100~nm for the adiabatic fiber coupling approach. In both cases efficiency is limited by imperfect mode matching and does not degrade when cooled to cryogenic temperatures. We note that coupling efficiencies as high as 97~\% have been demonstrated with the adiabatic coupler approach when the fiber tapering profile is optimized~\cite{tiecke_efficient_2015}.

\subsection{Gluing}
\label{sectiongluing}
Following the bonding method established by McKenna et al. \cite{mckenna_cryogenic_2019}, a failure rate of $\sim$25~\% when cooling to mK temperatures is expected for angle polished fibers bonded to SOI grating couplers using Norland Optical Adhesive (NOA) 88. We replicate similar statistics in our own testing.  However, when bonding to chips with a thermally grown oxide or cladded with a HSQ layer (baked at 300~$^\circ$C) a significantly higher failure rate (upwards of 50~\%) was observed. In order to find a more reliable adhesive, rapid thermal cycling was performed with liquid nitrogen.

\begin{figure*}
    \centering
    \includegraphics[width=\textwidth]{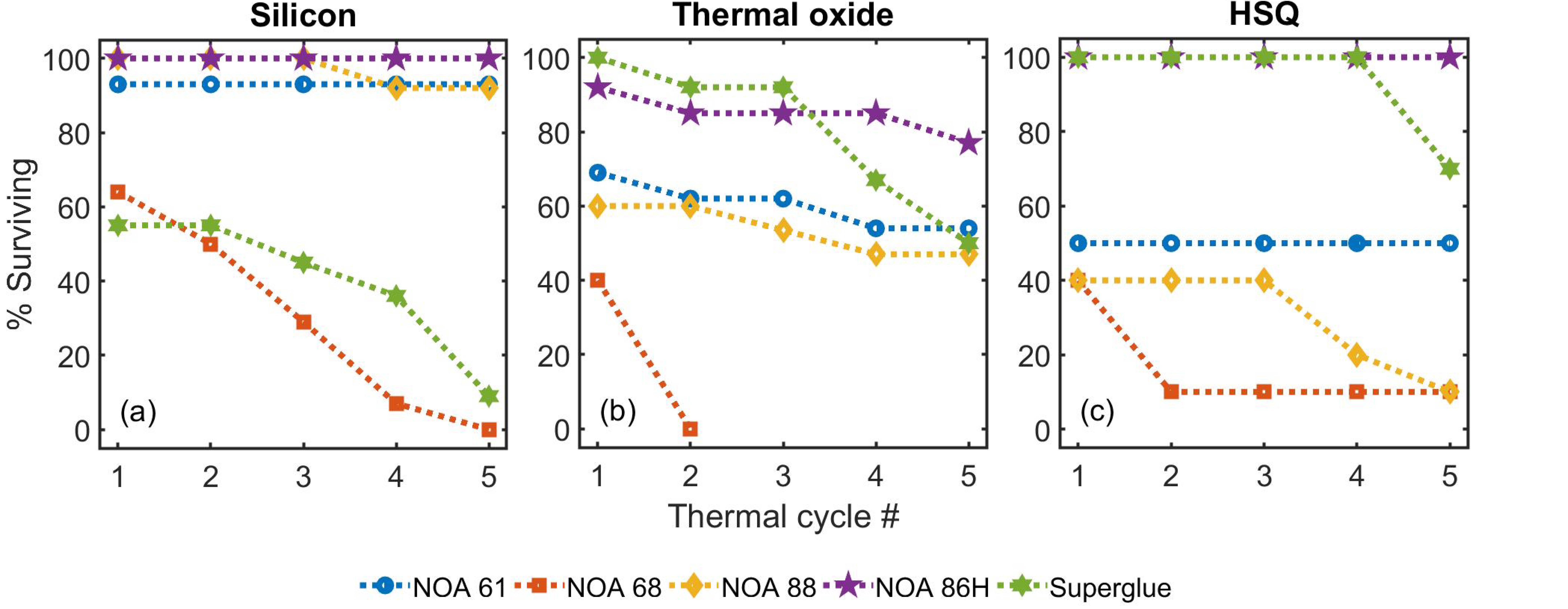}
    \caption{Various UV curing glues for optical bonding and superglue were tested for their durability against thermally cycling down to 77~K. Each thermal cycle corresponds to a cool-down from 300~K to 77~K followed by a warm-up back to 300~K. The left-most panel corresponds to  silicon as the bonding substrate, with N = 258 thermal cycles across all tested glues. The center panel employs a  substrate of thermally grown silica on a silicon wafer (Virginia Semiconductor) and represents N = 260 thermal cycles across all glues. Right panel: HSQ based solution spun onto silicon wafer and baked at 300~$^\circ$C to form a cladding, N = 172 thermal cycles across all the glues. In all tests the UV epoxy NOA 86H incorporating latent heat catalyst was found to yield the best bonding.} 
    \label{fig:StatsFigure}
\end{figure*}

Some common UV-curing glues were tested (NOA 88, NOA 68, and NOA 61) as well as several options containing a latent heat-curing catalyst. All UV glues with a latent heat curing catalyst tested (NOA 86H, NOA 86TLH, and NOA 89H) yielded similar results,  so only NOA 86H, which possesses the viscosity best suited for gluing fibers to a grating coupler or waveguide (250-350 centipoise) is included in Fig.~\ref{fig:StatsFigure}. Indeed, too high of a viscosity makes precise application difficult while too low allows the glue to spread far from the application point, potentially covering unwanted features. Cyanoacrylate glue (LOCTITE Super Glue) was also included for comparison because we have successfully used it for securing silica optical fiber tapers to a glass holder at mK temperatures in previous experiments~\cite{harris_laser_2016,mcauslan_microphotonic_2016,he_strong_2020}, however its low viscosity and long curing time make it impractical for bonding to grating couplers with micrometer precision.

To quickly test multiple glues on each substrate, we prepared sample chips of silicon, thermally grown silicon oxide, and HSQ cladding. These chips were attached to a brass block to reduce the temperature gradient during thermal cycles. We stripped and cleaned SMF-28 optical fiber before gluing it to the substrate following the gluing procedure from McKenna et al.\cite{mckenna_cryogenic_2019} and curing with a UV lamp (Polytec UV LC-5) for five minutes. Then NOA 61, 68, and 88 were baked for 12 hours at 42~$^\circ$C to age the glues and maximise their bonding strength. The glue containing a heat curing catalyst (NOA 86H) was cured in an oven at 125~$^\circ$C for 10 minutes and the superglue was left overnight to fully cure. The test chips were then slowly submerged in liquid nitrogen over 40 minutes and allowed to thermalize to 77~K before letting the liquid nitrogen boil off and the chips to naturally warm back to room temperature. Based on the expected thermal expansion down to 77~K, this simulates the majority of thermal stresses induced when cooling these devices down to 10~mK. With the harsher cooling rate, we find that bond survival during these liquid nitrogen thermal cycles is a good predictor of bond survival to mK temperatures.

The statistics for these tests is shown in Fig.~\ref{fig:StatsFigure}, where we plot the cumulative survival rate of fiber bonds to five successive thermal cycles. For each substrate material, we start with 15 fiber optic bonds for each glue, such that these statistics represent 690 discrete fiber bond stress tests overall. As noted above, the strength of the bond is strongly substrate dependent, with thermal oxide and HSQ providing much weaker bonds to NOA 88. Surprisingly NOA61 which is a common adhesive for bonding glass subject to temperature extremes was shown to work well on silicon substrates, but failed approximately 50~\% of the time on silica and HSQ substrates. In all cases the best performance was obtained through the use of NOA 86H. The mechanical stress transferred through the glue depends on the degree of polymerisation of the adhesive. UV glues typically achieve 85~\% polymerization with UV curing alone~\cite{mittal_progress_2018}, allowing stress to build up in the adhesive causing misalignment of the optical fiber or complete bond failure. Curing using a heat sensitive catalyst allows for nearly complete polymerization which greatly increases the bond strength, especially when cooling down to 10~mK. In particular, the latent heat catalyst fully cures even shadowed regions and overcomes issues potentially stemming from oxygen inhibition, which led to repeated bond failure using nanometer-thick layers of NOA 88 glue in the adiabatic fiber-coupling approach (illustrated in Fig.~\ref{fig:PhotonicDevice}~(b)) despite the waveguide material being silicon. Since identifying NOA 86H as an ideal candidate for both bond strength and ease of application, we have not yet observed failures in devices cooled to 10~mK in subsequent experiments.   

\section{Application to superfluid optomechanics} 
For superfluid optomechanics applications, good control over the achieved superfluid film thickness is required  to control the speed of sound~\cite{baker_theoretical_2016} and access different regimes, as depending on the thickness, the optomechanical interaction may be dominated by third sound waves, capillary waves (ripplons) or first sound~\cite{enss_low-temperature_2005, harris_laser_2016, kashkanova_superfluid_2017, he_strong_2020, harris_proposal_2020}.

\subsection{Superfluid film thickness}
\begin{figure*}
    \centering
    \includegraphics[width=\textwidth]{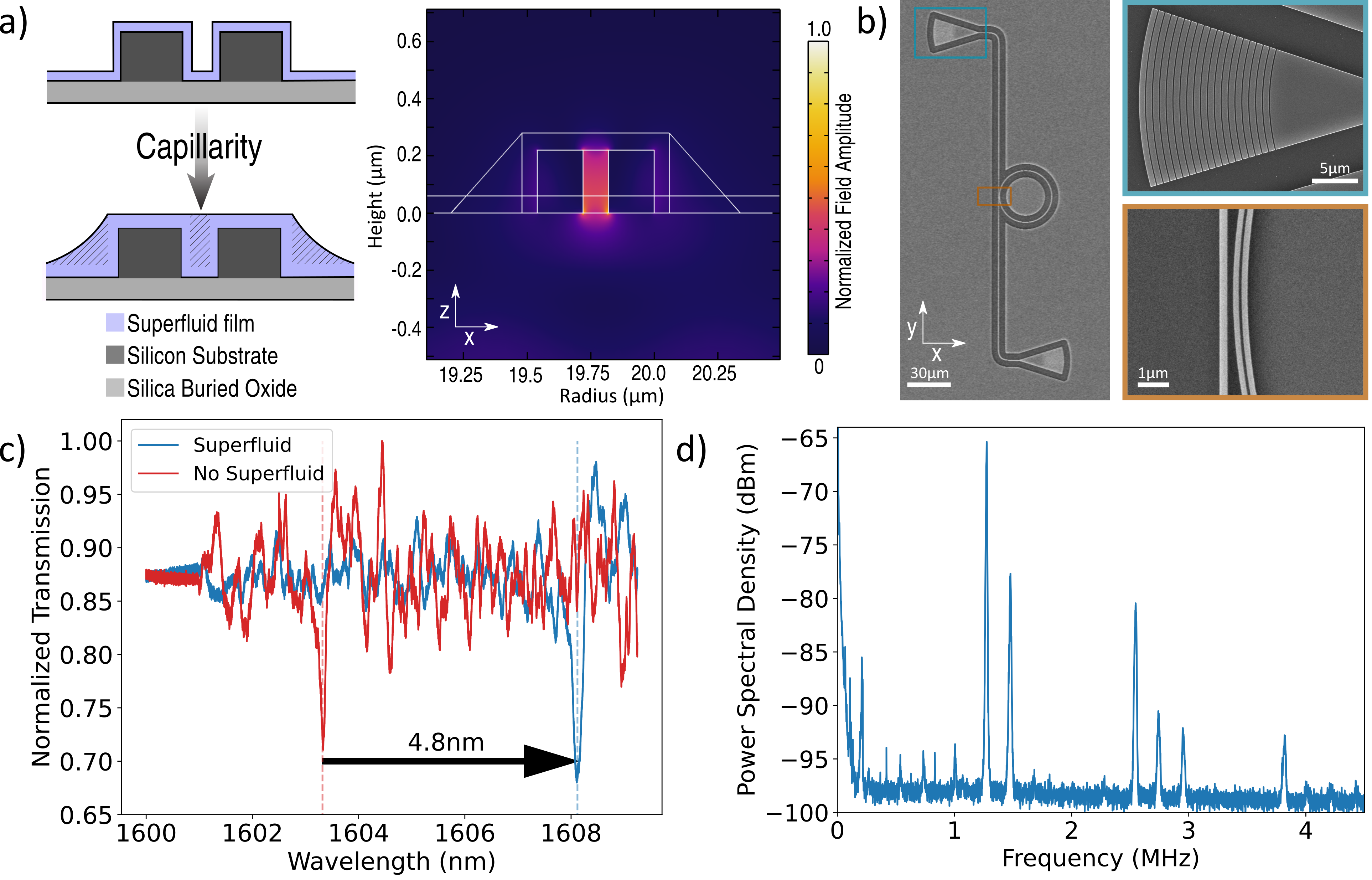}
    \caption{
    (a) The slot device proposed in \cite{harris_proposal_2020} is fabricated, and packaged with enough helium such that the capillarity fills larger volumes. (b) shows SEM images of the fabricated device, including zoom-ins on the grating coupler and slot resonator. (c) After packaging the device with helium and placing in a dilution refrigerator, the whispering gallery mode resonance (Q of 7300) shifts as expected by 4.8~nm due to the refractive index change when superfluid helium coats the device. (d) Optical power spectrum revealing MHz frequency superfluid acoustic modes.}
    \label{fig:Measurement}
\end{figure*}

The required room-temperature pressure of helium depends on the volume and surface area of the prepared puck, the positioning of the chip within it,  as well as the desired thickness of the superfluid film. There are two distinct regimes for evaluating the superfluid film thickness: the \textit{unsaturated vapour pressure regime} and the \textit{saturated vapour pressure regime}. The main distinction is the magnitude of the van der Waals potential in comparison to the gravitational potential---in the unsaturated vapour pressure regime the magnitude of the van der Waals potential $\mu_{vdW}$ is much greater than the magnitude of the gravitational potential $\mu_g$.

\paragraph{Unsaturated Vapour Pressure Regime ($\mu_{vdW}\gg \mu_g$)}
In this regime the superfluid will form in an approximately uniform thickness over all surfaces, depending only on the quantity of helium injected. For a sealed container made of a single material filled with helium-4 gas at temperature $T$ and pressure $P$, Eq. \ref{eq:unsat} gives the film thickness $d$. The relevant parameters of the container are the volume $V$ and surface area $A$. The remaining constants in the equation are the Boltzmann constant $k_B$, the mass of a helium-4 atom $m_{\mathrm{He}}$ and the superfluid helium density $\rho$:

\begin{align}
    d\simeq\frac{P V m_{\mathrm{He}}}{A \rho k_B T}
    \label{eq:unsat}
\end{align}

Small departures from Eq. (\ref{eq:unsat}) can arise due to differing van der Waals coefficients of the materials inside the puck~\cite{enss_low-temperature_2005}.

\paragraph{Saturated Vapour Pressure Regime ($\mu_{vdW}\sim \mu_g$)}
In this regime, the helium pressure in the chamber is equal to the saturated vapor pressure $p_0$, and any additional helium gas liquefies into a superfluid reservoir at the lowest point of the sample chamber. The film thickness $d$ is then determined by the height of the surface $z$ above the reservoir, which can be obtained by equating the van der Waals and gravitational chemical potentials \cite{enss_low-temperature_2005}  $\mu_{\mathrm{vdW}} = -\avdw ${\large{/}}$ d^3$ and $\mu_{\mathrm{grav}} = g z$:

\begin{align}
    d\left(z\right) &= \sqrt[3]{\frac{\avdw}{g z}}
    \label{eq:sat}
\end{align}

Here $\avdw$ is the van der Waals coefficient of the substrate ($=2.6 \times 10^{-24}\, \text{m}^5\text{s}^{-2}$ for silica~\cite{baker_theoretical_2016}), and $g=9.8$ m.s$^{-2}$ the gravitational acceleration. With the pinch-off tube facing downwards, the height $z$ between the chip and the reservoir level can be varied through the position of the crimp (see Fig.~\ref{fig:PuckSchematic}~(c)) between 1 and 10~cm,  which allows the film thickness to be precisely controlled over a range of 10 to 30~nm.  With the pinch-off tube facing upwards, the distance between the photonic device and the lowest point of the vessel can be reduced to the millimeter scale, resulting in a film thickness of up to 60~nm. Capillary action can be used to create regions with thicker volumes\cite{harris_proposal_2020}, as discussed below.

\subsection{Measurement}

A slot resonator as described in \cite{harris_proposal_2020} was used to validate the proposed packaging methods. For sufficient film thickness the slot of the resonator fills with superfluid helium through capillarity, as sketched in  Fig.~\ref{fig:Measurement}~(a), providing large overlap of the superfluid directly in the electrical field of the resonator. The slot resonator was fabricated on the top layer of a 15mm x 15mm silicon-on-insulator (SOI) (220~nm Si, 2~$\mu$m SiO$_2$, 500~$\mu$m Si) chip. The two annuli forming the slot resonator are 180~nm wide and are separated by a 100~nm slot, where the light will be predominantly confined, as shown in Fig.~\ref{fig:Measurement}~(a). The outer annulus has an outer radius of 20~$\mu$m. Fabrication dimensions and quality are confirmed with scanning electron microscope (SEM) images, shown in \ref{fig:Measurement}~(b)). Angle polished fibers (with an angle $\theta=35^{\circ}$) were aligned and glued using NOA 88 (prior to identifying NOA 86H) to the grating couplers, as described by McKenna et al.~\cite{mckenna_cryogenic_2019}. The device was then packaged as described above, encapsulating a pressure $P=1000$~mbar of helium-4 gas at room temperature.

The optical spectrum of the device was measured  at cryogenic temperatures both in the presence of superfluid and with no superfluid, as seen in \ref{fig:Measurement}~(c). From these spectra, the whispering gallery mode resonance wavelength shift due to superfluid is measured to be 4.8~nm, consistent with the filling of the slot with superfluid~\cite{harris_proposal_2020} and coverage by a film thickness of 55~nm. Using Eq. \ref{eq:sat} with the van der Waals coefficient for silica shows that this corresponds to the surface of the chip being 1.6 mm higher than the lowest point of the mounted package. The acoustic waves in the superfluid can be driven using a blue-detuned laser drive~\cite{harris_laser_2016}, using a combination of radiation pressure and fountain pressure forces. Their motion is transduced through a heterodyne optical measurement scheme~\cite{he_strong_2020}. Several acoustic modes are visible, as shown in \ref{fig:Measurement}~(d). Higher harmonics arise due to nonlinear optical transduction  caused by large amplitude superfluid motion, achieved here with microwatts of incident optical power. A full investigation of these or similar devices will be described elsewhere; these preliminary results serve to validate the optical integration, superfluid-tight packaging and film thickness control at mK temperatures. 

\section{Conclusion}
A cryogenic \emph{plug-and-play} packaging system for integrated photonic devices sealed with gas has been demonstrated. This allows for the use of standard cryostats and dilution refrigerators with optical fiber feedthroughs to be used for experiments requiring heat exchange gas or superfluid helium films. We systematically tested adhesives across common substrates for integrated photonics and discovered surprising variability in the failure rate depending on the substrate, and identified how the use of a heat-curing UV epoxy provides superior performance for these cryogenic applications. This enables  efficient fiber bonding to grating couplers with and without an upper oxide cladding, as well as the use of adiabatic fiber couplers for alignment-free broadband cryogenic optical coupling. While motivated here mainly by superfluid optomechanics, the applicability of the techniques developed here extend to emerging applications in quantum communications and control, such as optomechanical microwave to optical wavelength conversion~\cite{arnold_converting_2020,jiang_efficient_2020}, quantum optomechanics~\cite{riedinger_remote_2018,shomroni_optical_2019} and quantum photonics~\cite{elshaari_hybrid_2020}.

\begin{backmatter}

\bmsection{Funding} This work was funded by the US Army Research Office through grant number W911NF17-1-0310 and the Australian Research Council Centre of Excellence for Engineered Quantum Systems (EQUS, project number CE170100009). C.G.B acknowledges Australian Research Council Fellowship DE190100318 and UQ Grant UQECR2058737, G.I.H acknowledges Fellowship DE210100848.
\bmsection{Acknowledgments} The authors acknowledge the facilities, and the scientific and technical assistance, of the Microscopy Australia Facility at the Centre for Microscopy and Microanalysis (CMM), The University of Queensland.  This work was performed in part at the Queensland node of the Australian National Fabrication Facility, a company established under the National Collaborative Research Infrastructure Strategy to provide nano- and micro-fabrication facilities for Australia’s researchers.
\bmsection{Disclosures} The authors declare no conflicts of interest.
\bmsection{Data availability} Data underlying the results presented in this paper are not publicly available but may be obtained from the authors upon reasonable request.

\end{backmatter}



\bibliography{bibliography}
\end{document}